\documentclass[aps,pre,onecolumn,10pt,a4paper,superscriptaddress,nofootinbib]{revtex4-2}
\usepackage{microtype}
\usepackage{amsmath,amssymb,amsthm,mathtools}
\usepackage{booktabs}
\usepackage{graphicx}
\usepackage{xcolor,tikz}
\usetikzlibrary{arrows.meta}
\usepackage{hyperref}
\hypersetup{colorlinks=true,linkcolor=blue!45!black,citecolor=blue!45!black,urlcolor=blue!45!black}
\hypersetup{pdftitle={Solution-Space Geometry of Logarithmic-Width K-SAT and Self-Referential Hardness},pdfauthor={Chunyan Zhao},pdfsubject={Solution-space geometry of random K-SAT},pdfkeywords={K-SAT, solution independence, isolation, self-reference, Model RB}}
\usepackage{enumitem}
\setlist{nosep,leftmargin=*}
\newtheorem{theorem}{Theorem}[section]
\newtheorem{lemma}[theorem]{Lemma}
\newtheorem{proposition}[theorem]{Proposition}
\newtheorem{corollary}[theorem]{Corollary}
\theoremstyle{definition}
\newcommand{\E}{\mathbb E}
\newcommand{\Pp}{\mathbb P}
\newcommand{\Sol}{\mathcal S}

\newcommand{\TV}{d_{\mathrm{TV}}}
\newcommand{\Pois}{\operatorname{Poisson}}
\newcommand{\op}{o_{\mathbb P}}
\DeclareMathOperator{\Var}{Var}
\newcommand{\data}[1]{\section*{Data availability}#1}

\begin{document}
\title{Solution-Space Geometry of Logarithmic-Width $K$-SAT\\and Self-Referential Hardness}
\author{Chunyan Zhao}
\email{cyzhao@dase.ecnu.edu.cn}
\affiliation{School of Data Science and Engineering, East China Normal University, Shanghai, 200062, China}

\begin{abstract}
We study the solution space of random $K$-SAT with $K=\lceil(1+\varepsilon)\log_2N\rceil$, motivated by self-referential satisfiability. At normalized clause densities $1/(1+\varepsilon)<\theta<1$, almost every solution is isolated within a linear Hamming radius, although exponentially many adjacent solution pairs remain. Their endpoints have the same leading entropy density as the full solution set but carry a polynomially vanishing fraction of its uniform measure. In an explicit high-density interval, the typical nearest-neighbor distance divided by $N$ converges to $H_2^{-1}(\theta)$, where $H_2$ is binary entropy. The global minimum distance remains one at every fixed subcritical density. A shell second-moment calculation proves the distance law and determines the restricted entropy on its concentration region. The total entropy is smooth through local isolation. Local neighbors and global solutions disappear in Poisson windows whose widths differ by a factor of order $N/K$. In the critical window, distinct surviving solutions are separated by approximately $N/2$; the established Poisson count then gives unique-SAT/UNSAT coexistence. These results describe the witnesses used in self-referential blocking and distinguish the growing-width mechanism from fixed-width and locked constraints.
\end{abstract}
\maketitle
\begingroup
\makeatletter
\def\l@f@section{\addpenalty{\@secpenalty}\addvspace{.25em plus .5pt}}
\makeatother
\tableofcontents
\endgroup
\newpage
\section{Introduction}
Self-referential $K$-SAT is a random model for studying the structural origins of computational hardness. Its construction starts with a uniquely satisfiable formula, removes a redundant clause, and inserts a clause defined from the unique solution so as to exclude that solution. The resulting unsatisfiable companion preserves the unsigned clause--variable incidence graph. Fang et al.~\cite{FLL26} identify solution independence as a central feature supporting this construction: at an appropriate covering threshold, logarithmic-width random clauses leave a Poisson number of satisfying assignments, and both unique satisfiability and unsatisfiability have positive limiting probability.

A Poisson solution count specifies how many witnesses survive, but not how they are arranged. We study their entropy, Hamming distances, and weights under the uniform solution measure, using the statistical-mechanical description of random constraints~\cite{MM09}. The main question is whether a typical solution loses all nearby satisfying alternatives while the total number of solutions is still exponential. At the critical window, we also ask how this geometry supports unique satisfiability with nonvanishing probability. Throughout, unique-SAT/UNSAT coexistence means that both events have positive limiting probabilities under the same random law and parameter sequence.

\paragraph{The Model RB motivation.}
Xu and Li~\cite{XL00} introduced a growing-domain random constraint satisfaction model with an exact satisfiability threshold. Zhao, Zhou, Zheng, and Xu~\cite{ZZZX11} developed a message-passing treatment of growing-domain random constraints, connecting this ensemble to the cavity approach. Zhao, Zhang, Zheng, and Xu~\cite{ZZZX12} established clustering before the satisfiability threshold and investigated entropy using analytical and cavity-method calculations. Xu, Zhang, Liu, and Gong~\cite{XZLG15} subsequently proved a near-threshold organization into many separated small clusters and analyzed the absence of condensation in their parameter regime. Model RB also underlies the self-referential CSP construction of Xu and Zhou~\cite{XZ25}. These results motivate our analysis of local multiplicity and solution weights.

Model RB enlarges variable domains at fixed constraint arity. Here the domain remains Boolean and the clause width grows. Both models allow a direct comparison between solution counts and local freedom, with entropy normalized by their respective instance-size scales.

\paragraph{Constant-width geometry.}
The solution space of constant-width random $K$-SAT has been studied through clustering, freezing, and condensation. Achlioptas and Ricci-Tersenghi~\cite{ART09} proved frozen-variable structure in separated clusters in suitable fixed-width regimes. Krzaka\l a et al.~\cite{KMR07} and Montanari, Ricci-Tersenghi, and Semerjian~\cite{MRS08} developed the cavity description of cluster sizes and weights, distinguishing clustering from condensation. Ding, Sly, and Sun~\cite{DSS22} established the satisfiability threshold for sufficiently large fixed width using the corresponding replica-symmetry-breaking structure. Numerical overlap studies also resolve the organization of solutions as constraints are added~\cite{AAK07}. Freezing can be characterized through the size of rearrangements required to change a variable~\cite{S08}. The large-deviation analysis of the whitening process distinguishes frozen typical solutions from exceptional solutions with different frozen fractions~\cite{BDASZ16}. 

Locked constraint satisfaction problems provide a particularly close antecedent: their clustered phase consists of isolated solutions~\cite{ZM08}. Here the constraints are ordinary Boolean clauses, and isolation emerges as their width and density grow. A central feature is the simultaneous presence of macroscopically isolated typical solutions and exponentially many exceptional adjacent pairs.

\paragraph{Phase structure.}
Fix $\varepsilon>0$ and let $K=\lceil(1+\varepsilon)\log_2N\rceil$. Clauses are independent and uniform among clauses on $K$ distinct variables. Set
\[
 p=2^{-K},\quad q=1-p,\quad a=-\ln q,\quad M_c=\frac{N\ln2}{a}.
\]
For fixed $0<\theta<1$, we study $M=\lfloor\theta M_c\rfloor$. The two principal phenomena are:
\begin{enumerate}
\item \emph{Typical macroscopic isolation with exponentially many exceptional neighbors.} If $\theta>1/(1+\varepsilon)$, almost every solution has no distinct solution within distance $\delta N$ for every fixed $\delta<H_2^{-1}(\theta)$. Yet exponentially many adjacent pairs survive at each fixed $\theta<1$. Their endpoints have uniform solution mass $N^{1-\theta(1+\varepsilon)+o_{\mathbb P}(1)}$, which vanishes in the isolation regime. Their leading entropy density remains equal to that of the whole solution set, $(1-\theta)\ln2$.
\item \emph{A density-dependent typical distance alongside minimum distance one.} In the explicit high-density interval of Theorem~\ref{thm:nearest}, the nearest-neighbor distance of a uniform solution divided by $N$ converges to $H_2^{-1}(\theta)$. Throughout this interval the global minimum distance remains one with high probability. A typical witness and an extreme pair therefore probe different geometric scales of the same random formula.
\end{enumerate}
The entropy density remains linear through local isolation. The distance-shell calculation balances combinatorial multiplicity against survival under the constraints. We also compare the window in which local neighbors disappear with the later window in which the total solution count becomes finite.

The critical Poisson count and the blocking construction are inputs from~\cite{FLL26}. We combine that count with critical pair separation to describe a finite collection of witnesses at mutual distance $N/2+o(N)$, including empty and singleton outcomes with positive limiting probabilities. An elementary fixed-width solution-count gap provides the comparison at linear clause density. All statements about a typical solution refer to the uniform solution measure of the original random ensemble.

\paragraph{Physical interpretation.}
The total entropy and the closest pair do not determine a typical solution neighborhood. Exponentially many adjacent pairs can have vanishing equilibrium weight, leaving a uniform solution isolated on a linear distance scale. This is the geometric distinction that connects the counting results to self-referential hardness.

Section~\ref{sec:model} sets up the model and moment tools. Section~\ref{sec:isolation} proves macroscopic isolation. Section~\ref{sec:rare} studies rare close pairs and critical witnesses. Section~\ref{sec:comparison} compares these structures with Model RB and constant-width $K$-SAT. Section~\ref{sec:conclusion} summarizes the results and open questions.

\section{Random ensemble, observables, and solution independence}\label{sec:model}
\subsection{Solutions, local moves, and sampling}
A random formula $F$ is a conjunction of $M$ independently sampled clauses, each with $K$ distinct uniformly chosen variables and independent uniform signs. Repeated clauses occupy different indexed slots. Write $\Sol(F)\subseteq\{0,1\}^N$ for its solution set and $X=|\Sol(F)|$. Natural logarithms are denoted by $\ln$. Let
\[
 H(x)=-x\ln x-(1-x)\ln(1-x),\qquad H_2(x)=H(x)/\ln2.
\]
We use $0\ln0=0$ and $\ln0=-\infty$ when counting solutions. The inverse $H_2^{-1}$ is taken on $[0,1/2]$. Unless a different window is specified, a fixed density $\theta$ means $M=\lfloor\theta M_c\rfloor$. With high probability means with probability tending to one as $N\to\infty$, with $\varepsilon$ and each stated density parameter fixed.

For an integer $r\ge1$, let $G_r(F)$ have vertex set $\Sol(F)$, with two distinct vertices adjacent when their Hamming distance is at most $r$. For $s\in\Sol(F)$, define
\[
 D_F(s)=|\{t\in\Sol(F):d_H(s,t)=1\}|,\qquad
 R_F(s)=\min_{t\in\Sol(F)\setminus\{s\}}d_H(s,t),
\]
where $R_F(s)=\infty$ when $s$ is the unique solution. The uniform solution measure is $\pi_F(s)=1/X$ on satisfiable formulas. Fractions of solutions are assigned value zero on unsatisfiable formulas when needed.

On satisfiable samples, $\pi_F$ is the uniform zero-temperature measure. Its entropy density and the spin overlap of two assignments are
\begin{equation}\label{eq:observables}
 s_N(F)=\frac{\ln X}{N},\qquad Q(s,t)=1-\frac{2d_H(s,t)}{N}.
\end{equation}
The degree $D_F(s)$ measures single-flip freedom; $R_F(s)$ measures distance to the next solution. Components of $G_r(F)$ describe connectivity at move radius $r$.

We use two distributions on formula--solution pairs. Under $\mathsf P^\star$, choose $s$ uniformly and sample $F$ conditional on $s\in\Sol(F)$. Under $\mathsf U$, sample $F$ from the original distribution and then choose a uniform solution; if $X=0$, use a distinguished symbol. The transfer between these distributions is proved below.

\subsection{Counting and collision estimates}
The first moment is exact:
\begin{equation}\label{eq:first}
 \mu:=\E X=2^Nq^M.
\end{equation}
For assignments at Hamming distance $h$, put
\[
 \rho_N(h)=\frac{\binom{N-h}{K}}{\binom NK},\qquad
 A_h=1-2p+p\rho_N(h),
\]
with a zero numerator when $N-h<K$. A clause is falsified by both assignments with probability $p\rho_N(h)$. Thus
\begin{equation}\label{eq:second}
 \frac{\E[X(X-1)]}{\mu^2}
 =2^{-N}\sum_{h=1}^N\binom Nh\left(\frac{A_h}{q^2}\right)^M.
\end{equation}
Uniformly in $h$,
\begin{equation}\label{eq:expansion}
 \ln(A_h/q^2)=p\rho_N(h)+O(p^2).
\end{equation}
Here $Np=o(1)$ and $M_cp^2=o(1)$, so the accumulated error in~\eqref{eq:expansion} is $o(1)$ for every $M=O(M_c)$.

The following slightly flexible form of the collision estimate in~\cite{FLL26} is useful because parallel-edge counting replaces $(N,K)$ by $(N-1,K-1)$.
\begin{lemma}[Collision summability]\label{lem:collision}
Suppose $(1+\eta_0)\log_2n\le k\le C\log_2n$ for fixed $\eta_0>0$ and $C<\infty$. Put $\rho(h)=\binom{n-h}{k}/\binom nk$. For $B_n=n\ln2+O(1)$,
\[
 2^{-n}\sum_{h=1}^n\binom nh e^{B_n\rho(h)}=1+o(1).
\]
\end{lemma}
Appendix~\ref{app:collision} proves the lemma. Monotonicity in $B_n$ also gives an upper bound $1+o(1)$ whenever $0\le B_n\le n\ln2+C_0$ for a fixed constant $C_0$.

\begin{proposition}[Subcritical concentration]\label{prop:concentration}
For integer sequences $0\le M\le M_c$ with $\mu\to\infty$,
\[
 \E(X/\mu-1)^2\to0.
\]
In particular, for fixed $0<\theta<1$ and $M=\lfloor\theta M_c\rfloor$,
\begin{equation}\label{eq:entropy}
 X/\mu\xrightarrow{\mathbb P}1,\qquad
 \mu=2^{(1-\theta)N+o(1)},\qquad
 N^{-1}\log_2X\xrightarrow{\mathbb P}1-\theta.
\end{equation}
\end{proposition}
\begin{proof}
Since $p\le a$, we have $Mp\le N\ln2$ and $Mp^2=o(1)$. Equations~\eqref{eq:second}--\eqref{eq:expansion} and Lemma~\ref{lem:collision} bound the normalized factorial moment by $1+o(1)$. Hence
\[
 0\le\frac{\Var X}{\mu^2}
 =\frac{\E[X(X-1)]}{\mu^2}-1+\frac1\mu\le o(1)+\frac1\mu=o(1).
\]
Equation~\eqref{eq:first} gives the remaining assertions.
\end{proof}

\begin{lemma}[Transfer to a uniform solution]\label{lem:transfer}
Under the hypotheses of Proposition~\ref{prop:concentration},
\[
 \TV(\mathsf P^\star,\mathsf U)
 =\frac12\E\left|\frac X\mu-1\right|=o(1).
\]
\end{lemma}
\begin{proof}
The formula marginal under $\mathsf P^\star$ has density $X/\mu$ relative to the original formula law. Conditional on a satisfiable formula, both distributions select a uniform solution. Assigning the special symbol to unsatisfiable formulas gives the displayed total-variation identity, including their probability mass. Cauchy--Schwarz and Proposition~\ref{prop:concentration} finish the proof.
\end{proof}

The planted formula law is size-biased by $X$. Concentration of $X/\mu$ makes this bias vanish in total variation, allowing planted estimates to describe a uniform solution of the original formula law. Related planting arguments appear in locked constraint satisfaction~\cite{ZK11}. Equation~\eqref{eq:entropy} gives $s_N\to(1-\theta)\ln2$ in probability.

At the critical window we use the established Poisson theorem in its precise fixed-parameter form.
\begin{theorem}[Critical count, Fang et al.~\cite{FLL26}]\label{thm:poisson}
For every fixed $t\in\mathbb R$, set
\[
 M_N(t)=\left\lfloor\frac{N\ln2+t}{a}\right\rfloor.
\]
Then $\TV(\mathcal L(X),\Pois(e^{-t}))\to0$.
\end{theorem}

The corresponding factorial second moment satisfies $\E[X(X-1)]/\mu^2\to1$: equations~\eqref{eq:second}--\eqref{eq:expansion} give a collision sum with $Mp=N\ln2+t+o(1)$, and Lemma~\ref{lem:collision} applies.

\section{Macroscopic isolation of typical solutions}\label{sec:isolation}
\subsection{Disappearance of single-coordinate freedom}
Let
\begin{equation}\label{eq:b}
 b=\frac{Kp}{Nq},\qquad d_N=N(1-b)^M.
\end{equation}
\begin{lemma}[Neighbor factorial moments]\label{lem:neighbors}
For a fixed assignment $s$, conditional on $s\in\Sol(F)$, and every integer $0\le r\le N$,
\begin{equation}\label{eq:neighbor-moment}
 \E[(D_F(s))_r]=(N)_r(1-rb)^M.
\end{equation}
In particular, $\E D_F(s)=d_N$ and $\Var D_F(s)\le d_N$.
\end{lemma}
\begin{proof}
Flipping coordinate $i$ destroys a satisfied clause precisely when the literal on $i$ is its unique true literal under $s$. The unconditional probability of this event is $(K/N)p$; conditional on satisfaction it is $b$. A clause cannot have two different unique true literals. Thus it preserves all $r$ specified single-coordinate neighbors with conditional probability $1-rb$. Conditional clauses remain independent, giving~\eqref{eq:neighbor-moment}. Finally, $(N)_2(1-2b)^M\le N^2(1-b)^{2M}$ implies the variance bound.
\end{proof}

\begin{proposition}[Local isolation threshold]\label{prop:local}
Let $0<\theta<1$ be fixed and $M=\lfloor\theta M_c\rfloor$. Then
\[
 d_N=N^{1-\theta(1+\varepsilon)+o(1)}.
\]
Under $\mathsf U$, if $\theta<1/(1+\varepsilon)$ then $D_F(s)/d_N\to1$ in probability; if $\theta>1/(1+\varepsilon)$ then $\Pp(D_F(s)=0)\to1$.
\end{proposition}
\begin{proof}
From $M=\theta N\ln2/a+O(1)$ and $p/(qa)=1+O(p)$,
\[
 Mb=\theta K\ln2+o(1),\qquad Mb^2=o(1).
\]
This proves the expression for $d_N$. Apply Chebyshev's inequality when $d_N\to\infty$ and Markov's inequality when $d_N\to0$, then use Lemma~\ref{lem:transfer}.
\end{proof}

The relevant threshold concerns the neighborhood of a typical solution. A stronger statement follows by excluding all other solutions in a linear-radius ball.

\subsection{A linear isolation radius}
\begin{theorem}[Macroscopic isolation]\label{thm:macro}
Fix $1/(1+\varepsilon)<\theta<1$ and $0<\delta<H_2^{-1}(\theta)$. With $M=\lfloor\theta M_c\rfloor$,
\begin{equation}\label{eq:macro}
 \frac1X\bigl|\{s\in\Sol(F):R_F(s)>\delta N\}\bigr|
 \xrightarrow{\mathbb P}1.
\end{equation}
Simultaneously, $N^{-1}\log_2X\to1-\theta$ in probability.
\end{theorem}
\begin{proof}
First work conditional on a fixed satisfying assignment $s$. Writing conditional expectation as $\E_s$, the number $Z_h(s)$ of solutions at distance $h$ from $s$ satisfies
\begin{equation}\label{eq:shell}
 \E_s Z_h(s)=\binom Nh
 \left(1-\frac{p(1-\rho_N(h))}{q}\right)^M
 \le\binom Nh\exp\left\{-\frac{Mp}{q}(1-\rho_N(h))\right\}.
\end{equation}
Since $Np=o(1)$,
\begin{equation}\label{eq:coefficient}
 Mp/q=\theta N\ln2+o(1).
\end{equation}
Choose $\eta>0$ small enough that
$\theta(1+\varepsilon)(1-\eta/2)>1$.
For $1\le h\le\eta N/K$,
\[
 1-\rho_N(h)\ge1-e^{-Kh/N}\ge(1-\eta/2)Kh/N.
\]
For some fixed $\gamma>0$, the contribution of this range is bounded by
\[
 \sum_{h\ge1}\frac{N^h}{h!}N^{-(1+\gamma)h}=o(1).
\]
Choose a fixed $\beta\in(0,\delta)$ so small that
$H(\beta)<\theta\ln2(1-e^{-\eta})$.
For $\eta N/K<h\le\beta N$, we have $\rho_N(h)\le e^{-\eta}$, so the total contribution to~\eqref{eq:shell} is at most
\[
 (N+1)\exp\{NH(\beta)-\theta N\ln2(1-e^{-\eta})+o(N)\}
 =e^{-\Omega(N)}.
\]
For $\beta N<h\le\delta N$, the bound $\rho_N(h)\le(1-\beta)^K=o(1)$ yields a total contribution at most
\[
 (N+1)\exp\{N[H(\delta)-\theta\ln2]+o(N)\}
 =e^{-\Omega(N)}.
\]
Thus $\sum_{h=1}^{\lfloor\delta N\rfloor}\E_s Z_h(s)=o(1)$. Markov's inequality proves that the conditional probability of any other solution in this ball tends to zero. Lemma~\ref{lem:transfer} gives the same statement under $\mathsf U$.

Let $Q(F)$ be the fraction of solutions with a distinct neighbor within distance $\delta N$, set to zero if $X=0$. The preceding probability is $\E Q(F)$, up to the vanishing unsatisfiable event. Hence $Q(F)\to0$ in probability by Markov's inequality. Proposition~\ref{prop:concentration} supplies the entropy assertion.
\end{proof}

\begin{corollary}[Robust singleton mass]\label{cor:singleton}
Under the hypotheses of Theorem~\ref{thm:macro}, let $r_N\ge1$ be any deterministic integer sequence with $r_N=o(N)$. The fraction of solutions that are singleton components of $G_{r_N}(F)$ tends to one in probability. For every deterministic $L_N=2^{o(N)}$, the maximum $\pi_F$-mass of a union of at most $L_N$ components of $G_{r_N}(F)$ tends to zero in probability.
\end{corollary}
\begin{proof}
For large $N$, $r_N<\delta N$. Every solution counted in~\eqref{eq:macro} has no incident edge in $G_{r_N}(F)$. If $Q_N(F)$ is the mass of nonsingleton components, then $Q_N(F)\to0$ in probability. Any union of at most $L_N$ components has mass at most $Q_N(F)+L_N/X$, which tends to zero by~\eqref{eq:entropy}.
\end{proof}

Thus allowing any sublinear number of simultaneous flips leaves almost every solution in a singleton component.

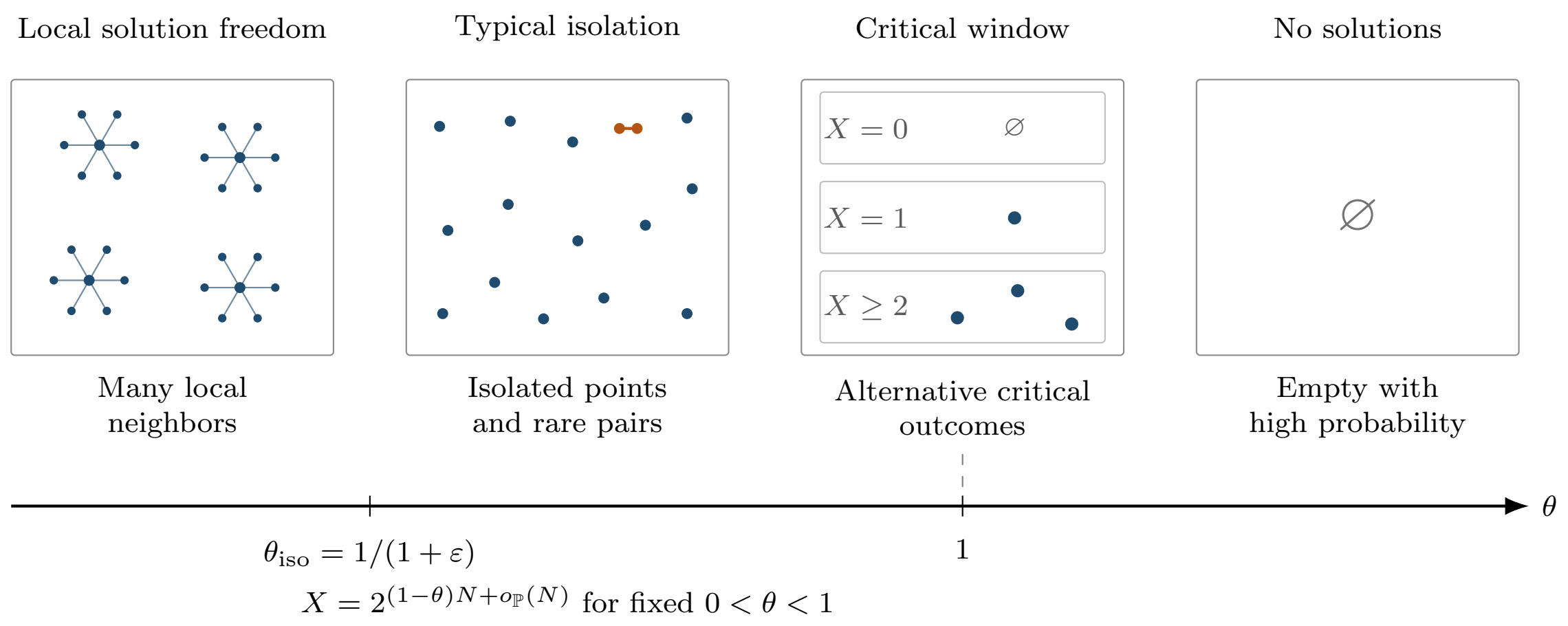
\begin{figure}[tb]
\centering
\begin{tikzpicture}[x=1cm,y=1cm,font=\footnotesize]
\definecolor{solblue}{RGB}{30,75,110}
\definecolor{pairorange}{RGB}{180,85,20}
\foreach \x in {0,3.8,7.6,11.4}{
 \draw[black!45,rounded corners=1pt] (\x,0) rectangle +(3.1,2.65);
}
\node[align=center] at (1.55,3.15) {Local solution freedom};
\node[align=center] at (5.35,3.15) {Typical isolation};
\node[align=center] at (9.15,3.15) {Critical window};
\node[align=center] at (12.95,3.15) {No solutions};
\foreach \cx/\cy in {.75/.72,2.2/1.9,.85/2.02,2.2/.65}{
 \foreach \ang in {0,60,120,180,240,300}{
  \draw[solblue!65,line width=.4pt] (\cx,\cy)--++(\ang:.34);
  \fill[solblue] (\cx,\cy) ++(\ang:.34) circle (1.15pt);
 }
 \fill[solblue] (\cx,\cy) circle (1.5pt);
}
\foreach \x/\y in {4.15/.4,4.65/.7,5.12/.35,5.7/.55,6.5/.4,4.2/1.2,4.78/1.45,5.45/1.1,6.1/1.25,6.55/1.6,4.12/2.2,4.8/2.25,5.4/2.05,6.5/2.28}{
 \fill[solblue] (\x,\y) circle (1.5pt);
}
\draw[pairorange,thick] (5.85,2.18)--(6.02,2.18);
\fill[pairorange] (5.85,2.18) circle (1.5pt);
\fill[pairorange] (6.02,2.18) circle (1.5pt);
\foreach \y in {.12,.98,1.84}{
 \draw[black!25,rounded corners=1pt] (7.78,\y) rectangle +(2.74,.69);
}
\node[black!65] at (8.22,2.18) {$X=0$};
\node[black!65] at (9.65,2.18) {$\varnothing$};
\node[black!65] at (8.22,1.32) {$X=1$};
\fill[solblue] (9.65,1.32) circle (1.8pt);
\node[black!65] at (8.22,.46) {$X\ge2$};
\foreach \x/\y in {9.1/.36,9.68/.62,10.2/.3}{\fill[solblue] (\x,\y) circle (1.8pt);}
\node[black!55,font=\Large] at (12.95,1.33) {$\varnothing$};
\node[align=center] at (1.55,-.5) {Many local\\neighbors};
\node[align=center] at (5.35,-.5) {Isolated points\\and rare pairs};
\node[align=center] at (9.15,-.5) {Alternative critical\\outcomes};
\node[align=center] at (12.95,-.5) {Empty with\\high probability};
\draw[-{Latex},thick] (0,-1.45)--(14.6,-1.45) node[right] {$\theta$};
\draw (3.45,-1.35)--(3.45,-1.55) node[below=3pt] {$\theta_{\mathrm{iso}}=1/(1+\varepsilon)$};
\draw (9.15,-1.35)--(9.15,-1.55) node[below=3pt] {$1$};
\draw[dashed,black!50] (9.15,-.95)--(9.15,-1.3);
\node[align=center] at (5.35,-2.35) {$X=2^{(1-\theta)N+o_{\mathbb P}(N)}$ for fixed $0<\theta<1$};
\end{tikzpicture}
\caption{Schematic solution-space evolution with increasing clause density, using the sequence-of-snapshots representation of figure~2 in~\cite{KMR07}. Dots represent solutions and links represent single-coordinate satisfying moves. Below $\theta_{\mathrm{iso}}$, local stars illustrate the neighbors of typical solutions. Above it, almost all solution mass lies on macroscopically isolated points, while exponentially many adjacent pairs persist with vanishing total mass at each fixed $\theta<1$. The three boxes within the critical panel show alternative outcomes of the same ensemble: zero, one, or several separated solutions. Their probabilities are specified by Corollary~\ref{cor:coexistence}. For fixed $\theta>1$, $\E X=2^{(1-\theta)N+o(1)}\to0$, giving an empty set with high probability. The critical window has relative width $\Theta(N^{-1})$. Panel positions, point counts, and drawn distances are schematic; the first panel specifies local neighborhoods rather than global connectivity.}
\label{fig:regimes}
\end{figure}

\subsection{A matching nearest-neighbor law}
The isolation radius has a matching upper bound in a fixed-density interval below satisfiability. Define
\begin{equation}\label{eq:shell-boundary}
 x_\varepsilon=\frac{1-\sqrt{2^{\varepsilon/(1+\varepsilon)}-1}}2,
 \qquad \theta_\varepsilon=H_2(x_\varepsilon),\qquad
 r_\theta=H_2^{-1}(\theta).
\end{equation}
The constant $x_\varepsilon$ is characterized by
$(1+\varepsilon)[-\log_2(1-2x_\varepsilon(1-x_\varepsilon))]=1$.
It specifies a sufficient decorrelation condition for the shell second moment below.

\begin{lemma}[Concentration on a distance shell]\label{lem:shell-concentration}
Take $M=\lfloor\theta M_c\rfloor$, where $0<\theta<1$, and fix $0<x<1/2$ such that
\[
 H_2(x)>\theta,\qquad
 (1+\varepsilon)[-\log_2(1-2x(1-x))]>1.
\]
For a fixed satisfying reference assignment $s$, let $Z_x(s)$ count solutions at distance $h=\lfloor xN\rfloor$. Under the conditional formula law given $s\in\Sol(F)$,
\[
 \frac{Z_x(s)}{\E[Z_x(s)\mid s\in\Sol(F)]}\xrightarrow{\mathbb P}1,
 \qquad
 \ln\E[Z_x(s)\mid s\in\Sol(F)]
 =N[H(x)-\theta\ln2]+o(N).
\]
\end{lemma}
\begin{proof}
Write $L=\binom Nh$, $r=\rho_N(h)$ and $v=(1-2p+pr)/q$. The mean is $Lv^M$, with the stated exponential rate by~\eqref{eq:shell}. For two assignments $t,u$ on the shell, write $d=d_H(t,u)$ and let $p\tau$ be the probability that one clause is falsified by all three of $s,t,u$. Inclusion--exclusion gives their joint conditional survival probability per clause as
\[
 w_{t,u}=\frac{1-3p+p(2r+\rho_N(d))-p\tau}{q}.
\]
Since $0\le\tau\le1$, uniformly over these pairs (when $t=u$, $d=0$ and $\tau=r$),
\[
 \ln(w_{t,u}/v^2)
 =p[\rho_N(d)-\tau]+O(p^2)
 \le p\rho_N(d)+O(p^2).
\]
For independent uniform assignments $T,U$ on the shell, put $D=d_H(T,U)$. Since $Mp^2=o(1)$,
\begin{equation}\label{eq:shell-second}
 \frac{\E[Z_x(s)^2\mid s\in\Sol(F)]}
 {\E[Z_x(s)\mid s\in\Sol(F)]^2}
 \le e^{o(1)}\E\exp\{Mp\rho_N(D)\}.
\end{equation}
We show that the expectation on the right is $1+o(1)$.

Put $x_N=h/N$ and $d_*=2x(1-x)$. The shell-distance distribution is
\[
 \Pp(D=2j)=\frac{\binom hj\binom{N-h}j}{\binom Nh}.
\]
For $z=2j/N$, its logarithm equals
$N[g_{x_N}(z)-H(x_N)]+O(\ln N)$ uniformly on its support, where
\[
 g_y(z)=yH\!\left(\frac{z}{2y}\right)
       +(1-y)H\!\left(\frac{z}{2(1-y)}\right).
\]
On $0\le z\le2\min\{y,1-y\}$, this strictly concave function has its unique maximum $H(y)$ at $z=2y(1-y)$ and satisfies $g_y(0)=0$. Since $x_N\to x$, each strict entropy gap used below persists for $g_{x_N}$ for all sufficiently large $N$.
Choose $0<\beta<d_*/2$ small enough that
$\sup_{0\le z\le\beta}g_x(z)<H(x)-\theta\ln2$.
Using $Mp=\theta N\ln2+o(1)$ and $\rho_N\le1$, the contribution from $D\le\beta N$, including the diagonal $D=0$, is $e^{-\Omega(N)}$.

Choose $\zeta>0$ small enough that $\beta<d_*-\zeta$ and
$(1+\varepsilon)[-\log_2(1-d_*+\zeta)]>1$.
For $\beta N<D<(d_*-\zeta)N$, strict concavity gives probability $e^{-\Omega(N)}$, while
$Mp\rho_N(D)\le O(N)(1-\beta)^K=o(N)$.
This contribution also vanishes exponentially.
For $D\ge(d_*-\zeta)N$,
\[
 Mp\rho_N(D)\le O(N)(1-d_*+\zeta)^K=o(1),
\]
uniformly. Its contribution is at most $1+o(1)$. As the exponential is always at least one, the full expectation is $1+o(1)$. Equation~\eqref{eq:shell-second} gives a second-moment ratio at most $1+o(1)$. The ratio is at least one, so the relative variance tends to zero and Chebyshev's inequality proves concentration.
\end{proof}

\begin{theorem}[Typical nearest-neighbor distance]\label{thm:nearest}
Fix $1/(1+\varepsilon)<\theta<1$ and let
$u_\theta=\max\{r_\theta,x_\varepsilon\}$. For every $\eta>0$,
\[
 \frac1X\bigl|\{s\in\Sol(F):
 r_\theta-\eta<R_F(s)/N<u_\theta+\eta\}\bigr|
 \xrightarrow{\mathbb P}1.
\]
In particular, if also $\theta\ge\theta_\varepsilon$, then
\begin{equation}\label{eq:nearest-law}
 \frac1X\bigl|\{s\in\Sol(F):
 |R_F(s)/N-H_2^{-1}(\theta)|<\eta\}\bigr|
 \xrightarrow{\mathbb P}1.
\end{equation}
\end{theorem}
\begin{proof}
For $0<\eta<r_\theta$, the lower bound follows by choosing $r_\theta-\eta<\delta<r_\theta$ in Theorem~\ref{thm:macro}; for $\eta\ge r_\theta$ it is immediate. Choose
$u_\theta<x<\min\{u_\theta+\eta,1/2\}$.
Lemma~\ref{lem:shell-concentration} gives a solution at distance $\lfloor xN\rfloor$ from the planted reference with probability $1-o(1)$. Lemma~\ref{lem:transfer} transfers this event to $\mathsf U$; if $Q_x(F)$ is the fraction of solutions with no other solution at distance $\lfloor xN\rfloor$ (zero when $X=0$), this gives $\E Q_x(F)=o(1)$. Markov's inequality yields $Q_x(F)\to0$ in probability and hence the upper bound. If $\theta\ge\theta_\varepsilon$, monotonicity of $H_2$ gives $u_\theta=r_\theta$.
\end{proof}

The exact law quantifies how far a typical witness must move to reach another solution: the distance increases as $H_2^{-1}(\theta)$ and approaches $1/2$ as $\theta\uparrow1$, taking the large-$N$ limit first. For example, when $\varepsilon=1$, it applies for $\theta\ge\theta_\varepsilon\approx0.676$, whereas typical isolation starts at $\theta=1/2$. The boundary $\theta_\varepsilon$ is a sufficient condition for this proof, not an asserted geometric transition. Throughout this interval the global minimum distance still equals one, as shown next.

\section{Rare close pairs and the critical solution space}\label{sec:rare}
\subsection{Adjacent pairs survive below the threshold}
Theorem~\ref{thm:macro} concerns almost all solution mass. The global minimum distance records an extreme pair and behaves differently.

\begin{theorem}[Parallel solution edges]\label{thm:edges}
Fix $0<\theta<1$ and a prescribed coordinate $i$. Let $Y_i$ count cube edges in direction $i$ whose two endpoints satisfy $F$. Then
\[
 \nu_N:=\E Y_i=\frac\mu2(1-b)^M,\qquad
 Y_i/\nu_N\xrightarrow{\mathbb P}1,
\]
and
\[
 \ln\nu_N=(1-\theta)N\ln2-\theta K\ln2-\ln2+o(1).
\]
Consequently $N^{-1}\log_2Y_i\to1-\theta$ in probability and the minimum distance between distinct solutions is one with high probability.
\end{theorem}
\begin{proof}
A clause kills at least one endpoint of a fixed edge with probability
$u=p(1+K/N)$. Thus $\E Y_i=2^{N-1}(1-u)^M$; the identity $1-u=q(1-b)$ gives the mean and its asymptotic expression.

Identify the $2^{N-1}$ parallel edges with their assignments on the other coordinates. Two edges at distance $h$ in this cube are both hit by one clause with probability
\[
 v_h=p\left[\left(1-\frac KN\right)\rho_K(h)
       +\frac{2K}{N}\rho_{K-1}(h)\right],\quad
 \rho_j(h)=\frac{\binom{N-1-h}{j}}{\binom{N-1}{j}}.
\]
The two terms correspond to clauses excluding or including $i$. Since $\rho_K(h)\le\rho_{K-1}(h)$, $v_h\le u\rho_{K-1}(h)$. Therefore
\[
 \frac{\E[Y_i(Y_i-1)]}{\nu_N^2}
 =2^{-(N-1)}\sum_{h=1}^{N-1}\binom{N-1}{h}
 \left(\frac{1-2u+v_h}{(1-u)^2}\right)^M.
\]
The logarithm in the summand is $v_h+O(u^2)$ uniformly. Moreover $Mu^2=o(1)$ and $Mu=\theta N\ln2+O(\ln N)\le(N-1)\ln2$ for large $N$. For large $N$, $K-1\ge(1+\varepsilon/2)\log_2(N-1)$. The monotone upper bound following Lemma~\ref{lem:collision}, with $(n,k)=(N-1,K-1)$, therefore bounds this ratio by $1+o(1)$. Since $\nu_N\to\infty$, $\Var(Y_i)/\nu_N^2\to0$.
\end{proof}

\begin{proposition}[Vanishing mass of nonisolated solutions]\label{prop:rare}
If $1/(1+\varepsilon)<\theta<1$, and $B(F)$ is the number of solutions with at least one single-flip neighbor, then
\[
 B(F)=(1+\op(1))\mu d_N,\qquad
 \frac{B(F)}{X}=(1+\op(1))d_N,
 \quad d_N=N^{1-\theta(1+\varepsilon)+o(1)}.
\]
In particular, $B(F)$ is exponential while its fraction among all solutions vanishes.
\end{proposition}
\begin{proof}
Let $Y=\sum_{i=1}^NY_i$. Coordinate symmetry and Cauchy--Schwarz give
\[
 \Var Y\le\left(\sum_i\sqrt{\Var Y_i}\right)^2=o((\E Y)^2),
 \qquad 2\E Y=\mu d_N.
\]
Deterministically,
\[
 0\le2Y-B(F)\le T(F):=\sum_{s\in\Sol(F)}D_F(s)(D_F(s)-1).
\]
By Lemma~\ref{lem:neighbors},
\[
 \E T(F)=\mu(N)_2(1-2b)^M\le\mu d_N^2=o(\mu d_N).
\]
Markov's inequality, concentration of $Y$, and Proposition~\ref{prop:concentration} finish the proof.
\end{proof}

Adjacent-pair endpoints have the same leading entropy density as all solutions, but their relative weight decays polynomially. Theorem~\ref{thm:macro} shows that almost all remaining mass is macroscopically isolated.

\subsection{Critical separation and coexistence}
\begin{theorem}[Critical separation]\label{thm:critical}
Fix $t\in\mathbb R$ and take $M=M_N(t)$. For every fixed $\delta>0$,
\[
 \Pp\left(\exists s\ne s'\in\Sol(F):
 \left|d_H(s,s')/N-1/2\right|>\delta\right)\to0.
\]
The conclusion also holds conditional on $X\ge2$.
\end{theorem}
\begin{proof}
Here $\mu\to e^{-t}$ and $Mp=N\ln2+t+o(1)$. The expected number of ordered solution pairs at distance $h$ is
\[
 \mu^2 2^{-N}\binom Nh\exp\{Mp\rho_N(h)+o(1)\},
\]
with a uniform error. Put $d_N'=N^{-1/4}$. On $h\ge(1/2-d_N')N$,
\[
 \rho_N(h)\le(1/2+d_N')^K=p(1+o(1)),\qquad Mp\rho_N(h)=o(1).
\]
This region has binomial mass $1-o(1)$ and hence contributes $1+o(1)$ to the collision sum. Lemma~\ref{lem:collision} shows that the full sum is $1+o(1)$; its nonnegative lower-distance remainder is therefore $o(1)$. This covers $h<(1/2-\delta)N$. On $h>(1/2+\delta)N$, the exponential factor is $1+o(1)$ and the binomial upper tail is $o(1)$. Thus the expected number of exceptional pairs is $o(1)$, proving the result. Finally, Theorem~\ref{thm:poisson} gives $\Pp(X\ge2)\to1-(1+e^{-t})e^{-e^{-t}}>0$.
\end{proof}

Combining Theorems~\ref{thm:poisson} and~\ref{thm:critical} yields a finite set of mutually distant witnesses in the critical window. The global minimum distance one at fixed subcritical density gives way to separation near $N/2$ in this window, conditional on having at least two solutions. These statements use different density scales; the critical window is centered at $M_c$ and has width $\Theta(2^K)$.

\begin{corollary}[Critical coexistence]\label{cor:coexistence}
Fix $t\in\mathbb R$, take $M=M_N(t)$, and put $\lambda=e^{-t}$. Then
\begin{equation}\label{eq:coexistence}
 \Pp(X=0)\longrightarrow e^{-\lambda},\qquad
 \Pp(X=1)\longrightarrow\lambda e^{-\lambda}.
\end{equation}
For every fixed integer $r\ge2$ and every $\eta>0$,
\begin{equation}\label{eq:finite-geometry}
 \Pp\left(X=r,\ 
 \max_{\substack{s,s'\in\Sol(F)\\s\ne s'}}
 \left|\frac{d_H(s,s')}{N}-\frac12\right|<\eta\right)
 \longrightarrow e^{-\lambda}\frac{\lambda^r}{r!}.
\end{equation}
\end{corollary}
\begin{proof}
Theorem~\ref{thm:poisson} gives $\Pp(X=r)\to e^{-\lambda}\lambda^r/r!$ for every fixed $r\ge0$. For $r\ge2$, the difference between this probability and the probability in~\eqref{eq:finite-geometry} is bounded by the exceptional-pair probability in Theorem~\ref{thm:critical}, applied with $\delta=\eta/2$, and hence tends to zero.
\end{proof}

The critical ensemble thus supports an empty solution set, a single witness, and finite collections of widely separated witnesses. In particular, at $t=0$ both probabilities in~\eqref{eq:coexistence} converge to $e^{-1}$. The isolation results describe the disappearance of local satisfying alternatives before the count becomes finite; the Poisson law specifies the probabilities of the different critical counts. This distinction connects the geometry of individual solutions to coexistence in the random ensemble.

\subsection{The geometric setting of self-referential blocking}
The redundant-clause argument of~\cite{FLL26} supplies the unique-solution instances used in the blocking construction.
\begin{proposition}[Unique witnesses and support-preserving blocking]\label{prop:blocking}
For fixed $t$ and $M=M_N(t)$, with probability tending to $e^{-t}e^{-e^{-t}}$, the formula is uniquely satisfiable and has at least $M-N\ln N$ redundant clause slots. Every such formula admits an unsatisfiable companion obtained by replacing one redundant clause on the same variable set.
\end{proposition}
\begin{proof}
Let $W$ count assignments violating exactly one clause slot. Then
\[
 \E W=2^NMp q^{M-1}=\mu Mp/q=O(N).
\]
Hence $\Pp(W>N\ln N)=o(1)$. Here a clause slot is essential if deleting it changes the solution set. Such a slot has a witness satisfying all other slots and falsifying that slot; witnesses for different essential slots are distinct. There are at most $W$ essential slots. Theorem~\ref{thm:poisson} now gives the asserted probability.

Let $s$ be the unique solution and let $D$ be redundant, with variable set $J$. Replace $D$ by $B_s=\bigvee_{i\in J}\ell_i$, where $\ell_i=x_i$ if $s_i=0$ and $\ell_i=\neg x_i$ if $s_i=1$. Removing $D$ leaves $s$ as the unique solution, and $B_s$ excludes $s$. The replacement preserves $N,M,K$ and the unsigned incidence graph.
\end{proof}

The replacement uses the sole satisfying assignment to reverse satisfiability while keeping every clause support fixed.

\section{Discussion: geometry and self-referential hardness}\label{sec:comparison}
\subsection{Typical geometry, exceptional mass, and entropy}
The two main results distinguish statistics of a uniform solution from statistics of an extreme pair. Above $\theta_{\mathrm{iso}}=1/(1+\varepsilon)$, typical solutions are macroscopically isolated while exponentially many adjacent pairs survive. In the range of Theorem~\ref{thm:nearest}, the typical distance is $H_2^{-1}(\theta)N+o_{\mathbb P}(N)$, whereas the minimum distance is one. Corollary~\ref{cor:singleton} translates this separation into measure organization: singleton components carry almost all mass under any prescribed sublinear move radius, and any subexponential collection of components carries vanishing mass.

\paragraph{Equal leading entropy, unequal weight.}
Let $B(F)$ count solutions with a single-flip neighbor. Proposition~\ref{prop:rare} implies, throughout the isolation regime,
\[
 \frac{\ln B(F)}N,\ \frac{\ln X}N
 \xrightarrow{\mathbb P}(1-\theta)\ln2,
 \qquad
 \frac{B(F)}X=N^{1-\theta(1+\varepsilon)+o_{\mathbb P}(1)}\to0.
\]
The exceptional population and the full solution set have the same leading exponential growth rate, but different weights in the uniform measure. Keeping only the extensive entropy therefore misses the typical isolation transition. The polynomial correction determines whether a uniformly selected solution retains a local alternative. This gives a quantitative form of the distinction between typical and exceptional geometry also considered through whitening large deviations~\cite{BDASZ16}.

\paragraph{Smooth entropy through geometric isolation.}
The total entropy density $s(\theta)=(1-\theta)\ln2$ is smooth at $\theta_{\mathrm{iso}}$ and has value $\varepsilon\ln2/(1+\varepsilon)>0$ there. For fixed $\theta<\theta_{\mathrm{iso}}$, Proposition~\ref{prop:local} gives $R_F(s)=1$ with high probability under $\mathsf U$. For each fixed $\theta>\theta_{\mathrm{iso}}$, Theorem~\ref{thm:macro} gives $R_F(s)/N>\delta$ with high probability for every $\delta<H_2^{-1}(\theta)$. Taking $N\to\infty$ first, the typical nearest-neighbor scale thus changes from microscopic to extensive without a singularity in the total entropy. This is a geometric isolation transition within a positive-entropy phase.

\paragraph{Distance-shell entropy.}
For a satisfying reference $s$, define
\[
 \Sigma_N(x;s)=N^{-1}\ln Z_x(s),\qquad \ln0=-\infty.
\]
Equation~\eqref{eq:shell} gives, for fixed $0<x<1/2$,
\[
 N^{-1}\ln\E[Z_x(s)\mid s\in\Sol(F)]
 =H(x)-\theta\ln2+o(1).
\]
Here $H(x)$ is the combinatorial entropy of the distance shell and $\theta\ln2$ is the leading cost of surviving the random constraints. Under the hypotheses of Lemma~\ref{lem:shell-concentration}, concentration and Lemma~\ref{lem:transfer} give the typical-shell law
\[
 \Sigma_N(x;s)\xrightarrow{\mathbb P}H(x)-\theta\ln2
 \quad\text{under }\mathsf U.
\]
For fixed $0<x<H_2^{-1}(\theta)$ the negative mean exponent instead implies an empty shell with high probability; it does not describe a finite negative entropy of actual solutions. In the interval where the nearest-neighbor bounds match, $H(x)=\theta\ln2$ locates the onset of other satisfying witnesses around a typical reference. The shell balance thereby explains the density-dependent distance law, while the exceptional adjacent pairs retain minimum distance one elsewhere in the same solution set.

\begin{figure}[tb]
\centering
\includegraphics[width=\textwidth]{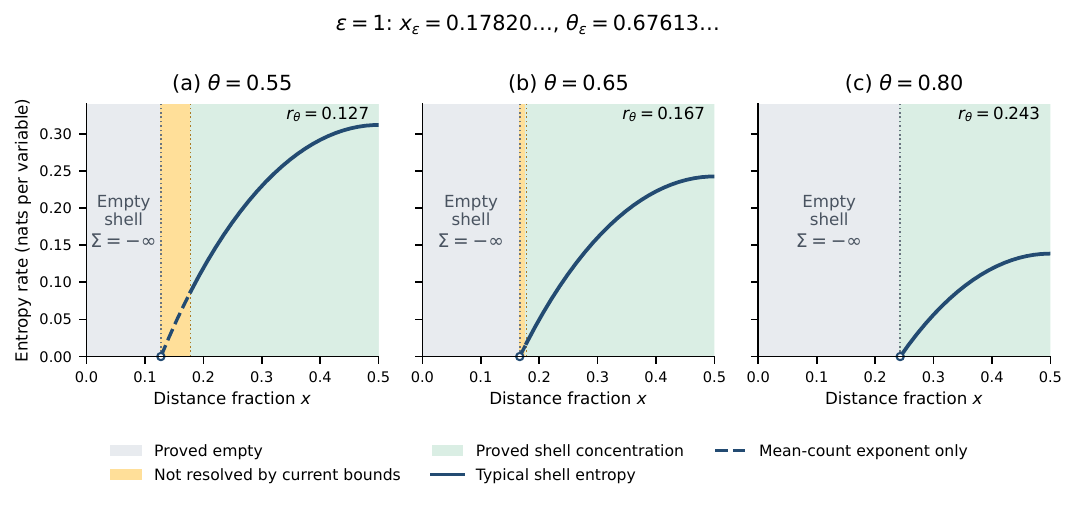}
\caption{Restricted entropy around a typical solution for $\varepsilon=1$ and $\theta=0.55,0.65,0.80$. Write $r_\theta=H_2^{-1}(\theta)$ and $x_\varepsilon=(1-\sqrt{\sqrt2-1})/2$. Gray: for each fixed $0<x<r_\theta$, the shell is empty with high probability and its actual entropy is $-\infty$; no finite entropy curve is drawn there. Theorem~\ref{thm:macro} also excludes all distinct solutions in balls of any fixed radius fraction below $r_\theta$ around almost every reference solution. Green: for each fixed $\max\{r_\theta,x_\varepsilon\}<x<1/2$, Lemma~\ref{lem:shell-concentration} proves shell-count concentration and the solid curve is the limiting typical entropy $H(x)-\theta\ln2$. Amber: the positive mean-count exponent is shown dashed, but the current bounds do not determine the typical shell entropy or establish nonemptiness. Panels (a) and (b) retain this band; panel (c) lies in the matching nearest-neighbor regime. The open zero markers indicate the mean-exponent roots, not a conclusion at equality. Shading specifies open asymptotic regions; boundary points, including $x=0$ and $x=1/2$, are not assigned by the figure. Negative mean exponents are omitted rather than plotted as finite negative solution entropies. Curves are analytical, not simulation data.}
\label{fig:restricted-entropy}
\end{figure}

Figure~\ref{fig:restricted-entropy} displays this distinction for three densities at $\varepsilon=1$. The unresolved band is $r_\theta<x\le x_\varepsilon$ when $r_\theta<x_\varepsilon$; it disappears from the plotted open intervals when $\theta>\theta_\varepsilon$. The dashed positive branch is a mean-count exponent, not a proved typical entropy. All regions refer to a uniformly sampled reference solution in the large-$N$ limit.

\subsection{Growing domains and growing width}
The RB results of Zhao et al.~\cite{ZZZX12} and Xu et al.~\cite{XZLG15} describe how growing domains produce separated solution groups. The logarithmic-width model instead produces singleton neighborhoods around almost every solution, together with a quantified exceptional population of adjacent pairs. In both comparisons the relevant quantity is the weight carried by a collection of local groups. Corollary~\ref{cor:singleton} shows that any subexponential collection carries vanishing mass here. The entropy scales differ: $N$ for the Boolean model and $N\ln N$ for growing-domain RB.

\subsection{Comparison with constant width}
Constant-width studies distinguish clustering, freezing, and condensation~\cite{ART09,KMR07,MRS08}. We compare these with local freedom and component weights under the uniform solution measure.

The standard fixed-width ensemble has a direct geometric obstruction to a finite nonempty solution set.
\begin{proposition}[Solution-count gap at constant width]\label{prop:fixed-gap}
Fix an integer $k\ge1$ and $\alpha>0$. Let $F$ have $M=\lfloor\alpha N\rfloor$ independent uniform clauses on $k$ distinct variables, and write $X=|\Sol(F)|$. For every $0<c<e^{-\alpha k}$,
\begin{equation}\label{eq:fixed-gap}
 \Pp(0<X<2^{cN})\longrightarrow0.
\end{equation}
In particular, $\Pp(X=1)\to0$.
\end{proposition}
\begin{proof}
Let $U$ count variables absent from every clause. For a specified variable the absence probability is $(1-k/N)^M$, so
\[
 \E U=N(1-k/N)^M=N e^{-\alpha k}+O(1).
\]
For two distinct variables the joint absence probability is
\[
 \left(\frac{\binom{N-2}{k}}{\binom Nk}\right)^M
 =\left(\frac{(N-k)(N-k-1)}{N(N-1)}\right)^M.
\]
Dividing it by $(1-k/N)^{2M}$ gives
\[
 \left(1-\frac{k}{(N-1)(N-k)}\right)^M=1+O(N^{-1}).
\]
Thus each off-diagonal covariance is $O(N^{-1})$ and $\Var U=O(N)$. Consequently $U/N\to e^{-\alpha k}$ in probability. On every satisfiable formula, arbitrary changes to the $U$ absent coordinates preserve satisfaction, and therefore $X\ge2^U$. Hence
\[
 \Pp(0<X<2^{cN})\le\Pp(U<cN)\longrightarrow0.
\]
For large $N$, $1<2^{cN}$, so the singleton assertion follows.
\end{proof}

The obstruction in Proposition~\ref{prop:fixed-gap} concerns the full standard ensemble. Its solution set factors exactly as $\Sol(F_{\mathrm{used}})\times\{0,1\}^U$, where $F_{\mathrm{used}}$ retains the variables that actually occur. Thus the unused coordinates alone exclude finite positive solution counts with nonvanishing probability. The induced formula on used variables has a different distribution; its internal geometry is a separate question.

On constrained variables, the fixed-width literature describes frozen coordinates within separated clusters~\cite{ART09}, the internal entropy and weights of pure states~\cite{MRS08}, and the rearrangements needed to change a variable~\cite{S08}. Removing absent coordinates does not identify these distinct observables with singleton components of $G_r(F)$. Our comparison therefore separates the elementary full-ensemble count obstruction from the literature on constrained-variable freezing and state entropy. In the logarithmic-width model, typical singleton dominance and the distance law are proved directly for the native uniform solution measure.

\begin{table}[tb]
\centering\small
\renewcommand{\arraystretch}{1.22}
\newcommand{\comparisonrow}[3]{\par\noindent
\begin{minipage}[t]{.18\textwidth}\raggedright #1\end{minipage}\hfill
\begin{minipage}[t]{.385\textwidth}\raggedright #2\end{minipage}\hfill
\begin{minipage}[t]{.385\textwidth}\raggedright #3\end{minipage}\par\smallskip}
\hrule\smallskip
\comparisonrow{Geometric question}{Fixed-width reference}{Logarithmic-width result}
\hrule\smallskip
\comparisonrow{Local freedom}{Full ensemble: unused coordinates give free subcubes. Constrained variables: cluster freezing is a separate observable.}{Typical solutions have no distinct solution within $\delta N$, for $\delta<H_2^{-1}(\theta)$.}
\comparisonrow{Mass organization}{Clustering and condensation distinguish the number and weights of pure states in the relevant regimes.}{Any $2^{o(N)}$ components under $o(N)$ moves carry vanishing mass in the macroscopic-isolation regime.}
\comparisonrow{Exceptional proximity}{Unused-coordinate subcubes supply adjacent solutions throughout the full solution set.}{Exponentially many adjacent pairs survive, but their endpoints have vanishing mass.}
\comparisonrow{Critical witnesses}{The fixed-width theory analyzes states and their internal entropy near satisfiability.}{Poisson total count; distinct critical witnesses lie at distance $N/2+o(N)$.}
\comparisonrow{Unique-SAT / UNSAT coexistence}{At fixed $k$ and linear clause density, with high probability $X=0$ or $X\ge 2^{cN}$ for some $c>0$; hence $\Pp(X=1)\to0$.}{In the critical window, $\Pp(X=0)\to e^{-\lambda}$ and $\Pp(X=1)\to\lambda e^{-\lambda}$, with $\lambda=e^{-t}$.}
\hrule
\caption{Comparison of structural observables. The logarithmic-width fixed-density statements assume $1/(1+\varepsilon)<\theta<1$; the last two rows use the critical window. Fixed-width cluster and pure-state statements refer to the regimes and definitions of the cited literature.}
\label{tab:comparison}
\end{table}

\subsection{Relation to locked constraint satisfaction}
In the locked occupation models of Zdeborov\'a and M\'ezard~\cite{ZM08}, allowed constraint occupancies are nonconsecutive and each variable belongs to at least two constraints. A single-variable flip therefore violates an incident constraint: adjacent satisfying assignments are excluded by the constraint rule. 

Ordinary disjunctive clauses allow local satisfying changes. Here increasing width and density suppresses those changes statistically around a typical solution, while exponentially many adjacent solution pairs remain at every fixed $\theta<1$. Above the isolation threshold their endpoints carry the explicitly quantified, polynomially vanishing mass of Proposition~\ref{prop:rare}. Isolation is therefore statistical: it holds for almost all solution mass, while the constraint rule still permits adjacent solutions. Theorem~\ref{thm:nearest} determines the typical distance on an explicit interval.

\subsection{Coexistence and self-referential blocking}
The local and global disappearance windows connect typical isolation to the self-referential construction. Appendix~\ref{app:window} gives a Poisson limit for the number of single-flip neighbors of a uniform reference solution; Theorem~\ref{thm:poisson} gives the established Poisson limit for the total solution count:
\[
 \begin{aligned}
 M=M_{\mathrm{iso}}(u):&\quad D_F(s)\Rightarrow\Pois(e^{-u}),
 &\Pr(D_F(s)=0)&\to e^{-e^{-u}},\\
 M=M_N(t):&\quad X\Rightarrow\Pois(e^{-t}),
 &\Pr(X=0)&\to e^{-e^{-t}}.
 \end{aligned}
\]
Their clause-count widths are respectively $\Theta(N2^K/K)$ and $\Theta(2^K)$, a ratio of order $N/K$. The first window concerns local alternatives around a surviving reference; the second concerns global witnesses. The local limit is a law for a uniformly sampled reference, and its zero probability gives the expected isolated fraction in that window.

At the global window, Corollary~\ref{cor:coexistence} gives isolated singleton witnesses and empty solution sets with positive limiting probabilities. Proposition~\ref{prop:fixed-gap} contrasts this with the full standard constant-width ensemble at linear density, where a surviving solution retains an extensive unused-coordinate subcube. Thus the geometry connects the loss of local freedom to the availability of a sole global witness.

Coexistence supplies unique-solution mother instances with nonvanishing probability. Proposition~\ref{prop:blocking} pairs each suitable mother with an unsatisfiable companion by replacing a redundant clause with one falsified by its sole witness. The random law supplies the mothers; the replacement defines their companions. The geometry describes the witnesses on which this construction acts: local alternatives have disappeared, and the remaining solutions are widely separated.

\section{Conclusion}\label{sec:conclusion}
Two geometric phenomena characterize the positive-entropy regime of logarithmic-width random $K$-SAT. Typical solutions are macroscopically isolated even though exponentially many adjacent pairs survive; the endpoints of those pairs have the same leading entropy density as the full solution set but only polynomially vanishing measure. In an explicit high-density interval, the typical nearest-neighbor distance follows $H_2^{-1}(\theta)N$, while the global minimum distance remains one at every fixed subcritical density. The closest pair therefore does not represent a typical solution neighborhood.

The total entropy stays smooth through local isolation, and distance-shell entropy identifies the balance controlling the typical distance law. Distinct Poisson windows describe the disappearance of local alternatives and of global witnesses. At the latter window, widely separated finite solution sets and positive singleton probability provide the geometric setting for self-referential blocking. Comparisons with Model RB, locked constraints, and fixed-width satisfiability identify the role of growing width in producing this organization. The remaining questions are to extend the exact nearest-neighbor law throughout the isolation regime, locate the finer density scale where exceptional adjacent pairs disappear, and understand the boundary $K\sim\log_2N$.

\data{This work is analytical. Proofs of the new results are included in the article and its appendices; the critical Poisson theorem is cited from~\cite{FLL26}. Figure~\ref{fig:restricted-entropy} is generated from the stated formulas; no simulation datasets were used.}

\clearpage\appendix
\section{Collision summability}\label{app:collision}
We prove Lemma~\ref{lem:collision} in the form needed for both assignment and parallel-edge counts. Write $p_k=2^{-k}$ and observe that
\[
 \rho(h)\le(1-h/n)^k\le e^{-kh/n}.
\]
Choose $\eta>0$ small enough that $(1+\eta_0)(1-\eta/2)>1$, and then choose $\gamma>0$ below the excess over one. Let $\beta\in(0,1/2)$ be small enough that
\[
 H(\beta)<\ln2(1-e^{-\eta}).
\]
Set $\xi=-(1+\eta_0)\log_2(1-\beta)>0$, choose $0<\zeta<\min\{1/2,\xi/2\}$, and put $d_n=n^{-\zeta}$.

For $1\le h\le\eta n/k$,
\[
 1-\rho(h)\ge(1-\eta/2)kh/n.
\]
Because $B_n=n\ln2+O(1)$, the corresponding weighted terms satisfy
\[
 2^{-n}e^{B_n\rho(h)}\le C_0 n^{-(1+\gamma)h}
\]
for some constant $C_0$. Their sum after multiplication by $\binom nh\le n^h/h!$ is $o(1)$.

For $\eta n/k<h\le\beta n$, $\rho(h)\le e^{-\eta}$, and the total contribution is at most
\[
 (n+1)\exp\{nH(\beta)-n\ln2+B_ne^{-\eta}\}=e^{-\Omega(n)}.
\]
For $\beta n<h<(1/2-d_n)n$, $\rho(h)\le(1-\beta)^k\le n^{-\xi}$. If $Z\sim\operatorname{Bin}(n,1/2)$, this contribution is bounded by
\[
 e^{O(n^{1-\xi})}\Pp(Z\le n/2-nd_n)
 \le\exp\{O(n^{1-\xi})-2n^{1-2\zeta}\}=o(1).
\]
Finally, for $h\ge(1/2-d_n)n$,
\[
 \rho(h)\le(1/2+d_n)^k=p_k(1+2d_n)^k=p_k(1+o(1)).
\]
Since $np_k=o(1)$, the exponential factor is uniformly $1+o(1)$. This range has binomial mass $1-o(1)$, so its contribution is $1+o(1)$. Adding the four ranges proves the lemma.

\section{The local isolation window}\label{app:window}
The neighbor count has an explicit transition window that complements the fixed-density statements.
\begin{proposition}
For fixed $u\in\mathbb R$, put
\[
 M_{\mathrm{iso}}(u)=\left\lfloor\frac{\ln N+u}{-\ln(1-b)}\right\rfloor,
 \qquad b=\frac{Kp}{Nq}.
\]
Under both $\mathsf P^\star$ and $\mathsf U$,
\[
 D_F(s)\Rightarrow\Pois(e^{-u}),\qquad
 \Pp(D_F(s)=0)\to e^{-e^{-u}}.
\]
Also $M_{\mathrm{iso}}(u)/M_c\to1/(1+\varepsilon)$, and the window scale is $\Theta(N2^K/K)$ clauses.
\end{proposition}
\begin{proof}
At this density, $Mb=\ln N+u+o(1)$ and $Mb^2=o(1)$. For each fixed $r$, Lemma~\ref{lem:neighbors} yields
\[
 \ln\E[(D_F(s))_r]
 =\ln(N)_r+M\ln(1-rb)=-ru+o(1).
\]
The factorial-moment criterion gives the Poisson limit under $\mathsf P^\star$. Since $M/M_c\to1/(1+\varepsilon)<1$, Proposition~\ref{prop:concentration} and Lemma~\ref{lem:transfer} apply, giving the limit under $\mathsf U$. The scale follows from $[-\ln(1-b)]^{-1}\sim b^{-1}$.
\end{proof}
The last probability is the limiting expected fraction of isolated solutions in this fine window. The fixed-density concentration statements in the main text follow from their separate first- and second-moment estimates.

\section{Doubleton components among exceptional solutions}\label{app:doubleton}
The exceptional nonisolated solutions have a simple dominant component type under single-coordinate moves.
\begin{proposition}
For fixed $1/(1+\varepsilon)<\theta<1$, the fraction of nonisolated solutions belonging to components of size two in $G_1(F)$ tends to one in probability.
\end{proposition}
\begin{proof}
Use $T(F)=\sum_sD_F(s)(D_F(s)-1)$ from the proof of Proposition~\ref{prop:rare}. Let $H$ be the vertices of degree at least two. Then
\[
 |H|\le T(F)/2,\qquad \sum_{s\in H}D_F(s)\le T(F).
\]
Every vertex in a component of size at least three either lies in $H$ or is a degree-one neighbor of a vertex in $H$. Thus there are at most $3T(F)/2$ such vertices. The proof of Proposition~\ref{prop:rare} gives $T(F)=\op(\mu d_N)$ and $B(F)=(1+\op(1))\mu d_N$. Since $B(F)>0$ with high probability, division by $B(F)$ proves the claim. The fraction may be set to zero on $B(F)=0$.
\end{proof}
\end{document}